\documentclass{PoS}

\usepackage{wrapfig}
\usepackage{graphicx}
\usepackage{xspace}

\newcommand{\GB}{\textsf{GAMBIT}\xspace}

\newcommand{\gambit}{\textsf{GAMBIT}\xspace}
\newcommand{\colliderbit}{\textsf{ColliderBit}\xspace}

\title{Global fits of the MSSM with GAMBIT}

\ShortTitle{Global fits of the MSSM with GAMBIT}

\author{\speaker{Anders Kvellestad}, {\rm on behalf of the GAMBIT collaboration}\\%
Department of Physics, University of Oslo, N-0316 Oslo, Norway\\
Department of Physics, Imperial College London, Blackett Laboratory, Prince Consort Road, London SW7 2AZ, UK\\
      E-mail: \email{anders.kvellestad@fys.uio.no}}

\abstract{The minimal supersymmetric standard model is a popular
  and well-motivated extension of the standard model.  As
  such, it has been constrained by a large number of different experimental searches.
  To truly assess the impacts of these experiments on the model one must
  perform a global fit, scanning over the multi-dimensional parameter
  space and combining all the data in a statistically rigorous
  manner.  In this talk, I presented results from global fits of
  supersymmetric models performed with \gambit, the Global and Modular Beyond-the-Standard-Model (BSM) Inference
  Tool.  I showed MSSM results from the latest
  \gambit papers, as well as exciting preliminary results from a dedicated
  study of the collider constraints on the electroweakino sector.}

\FullConference{39th International Conference on High Energy Physics\\
		  4-11 July 2018\\
		 Seoul, South Korea}

\begin{document}

Weak scale supersymmetric extensions of the standard model can solve
the large hierarchy problem, explain dark matter (DM), and achieve gauge coupling unification.

In recent years, the minimal supersymmetric standard model (MSSM) has
come under heavy pressure from a vast array of experiments in particle physics and
astrophysics.  To correctly assess their impact on the model, one must perform a
statistically rigorous combination of the data in a global fit. There
have been many global fits performed on the constrained MSSM (CMSSM). More recently, global fits have also been performed on weak scale parameterisations of the MSSM.

The Global and Modular BSM Inference Tool (\gambit) is both a state-of-the-art tool for performing global fits in essentially any
BSM theory, and a collaboration using it to perform global fits in a wide variety of BSM theories.  The \gambit code\footnote{\href{https://gambit.hepforge.org}{https://gambit.hepforge.org}} is public, and is accompanied by six component manuals \cite{gambit,ScannerBit,ColliderBit,FlavBit,DarkBit,SDPBit}.  Physics papers to date cover scalar singlet DM \cite{SSDM,SSDM2}, a weak-scale MSSM parameterisation with seven parameters \cite{MSSM} and the constrained, NUHM and NUHM2 variants of the MSSM \cite{CMSSM}.  Here, I present published and preliminary results from global fits of the MSSM, including an intriguing preliminary result in the electroweakino sector.

%To perform the global fits of the MSSM, \gambit uses a large number of external backend codes for efficient sampling (\Diver \cite{ScannerBit}, \MultiNest \cite{Feroz:2007kg,Feroz:2008xx}), mass spectra (\flexiblesusy \footnote{\flexiblesusy uses \sarah \cite{Staub:2009bi, Staub:2010jh,Staub:2012pb,Staub:2013tta} and contains some numerical routines from \softsusy \cite{Allanach:2001kg,Allanach:2013kza}.} \cite{Athron:2014yba, Athron:2016fuq, Athron:2017fvs}), decay BFs and widths (\HDECAY \cite{Djouadi:1997yw} and \SDECAY \cite{Muhlleitner:2003vg} in \SUSYHIT 1.5 \cite{Djouadi:2006bz}), Higgs likelihoods (\HiggsBounds and \HiggsSignals \cite{Bechtle:2008jh,HiggsSignals}); $(g-2)_\mu$ (\GMtwoCalc 1.3.1 \cite{gm2calc}), flavour physics (\SuperIso 3.6 \cite{Mahmoudi:2007vz}) and DM (\microOMEGAs 3.6.9.2 \cite{Belanger:2001fz}, \DarkSUSY 5.1.3 \cite{darksusy4}, \DDCalc 1.0.0 \cite{DarkBit}, \nuLike 1.0.4 \cite{IC22Methods,IC79_SUSY}).

\begin{figure*}[tbp]
  \centering
  \includegraphics[width=0.245\textwidth]{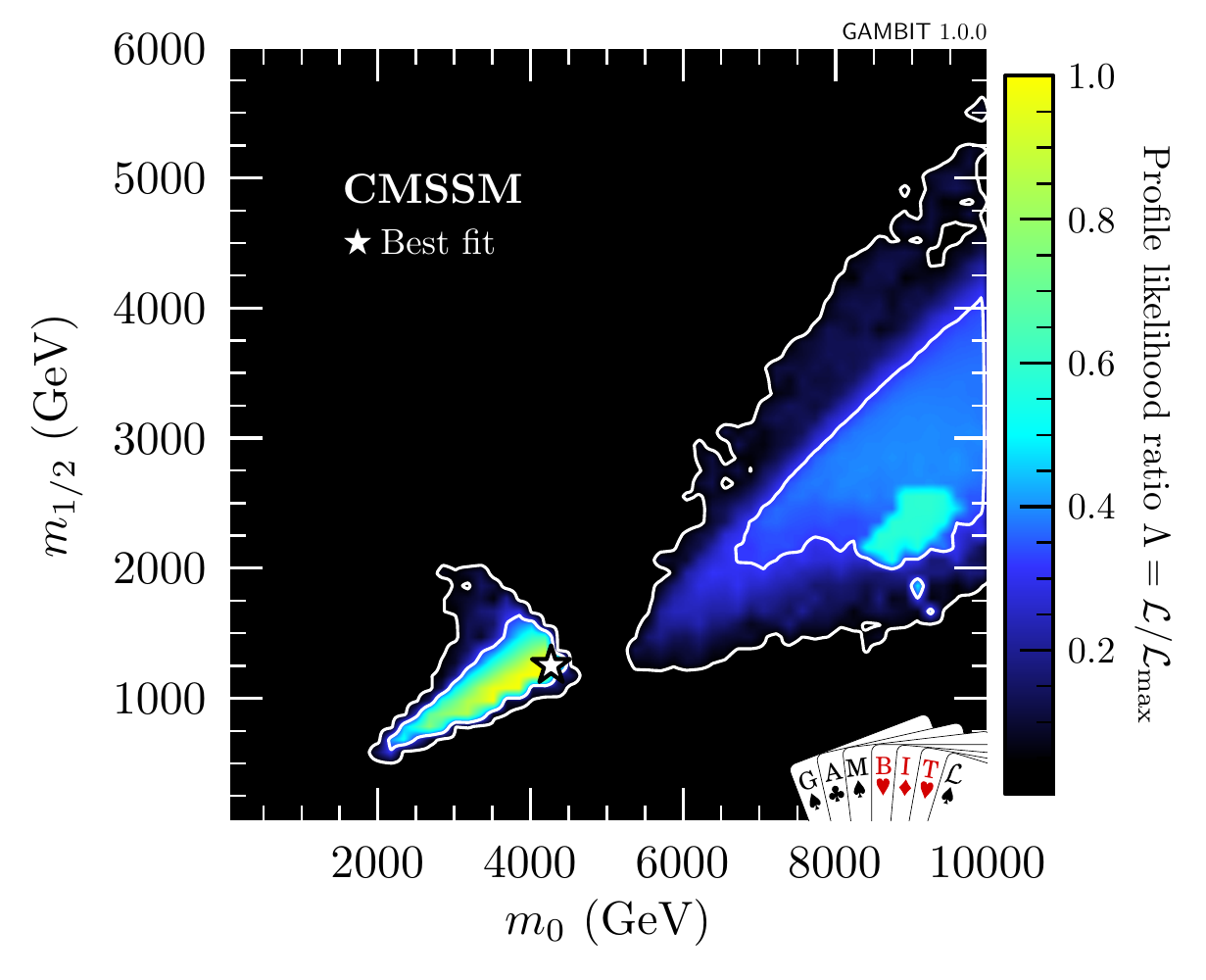}
  \includegraphics[width=0.245\textwidth]{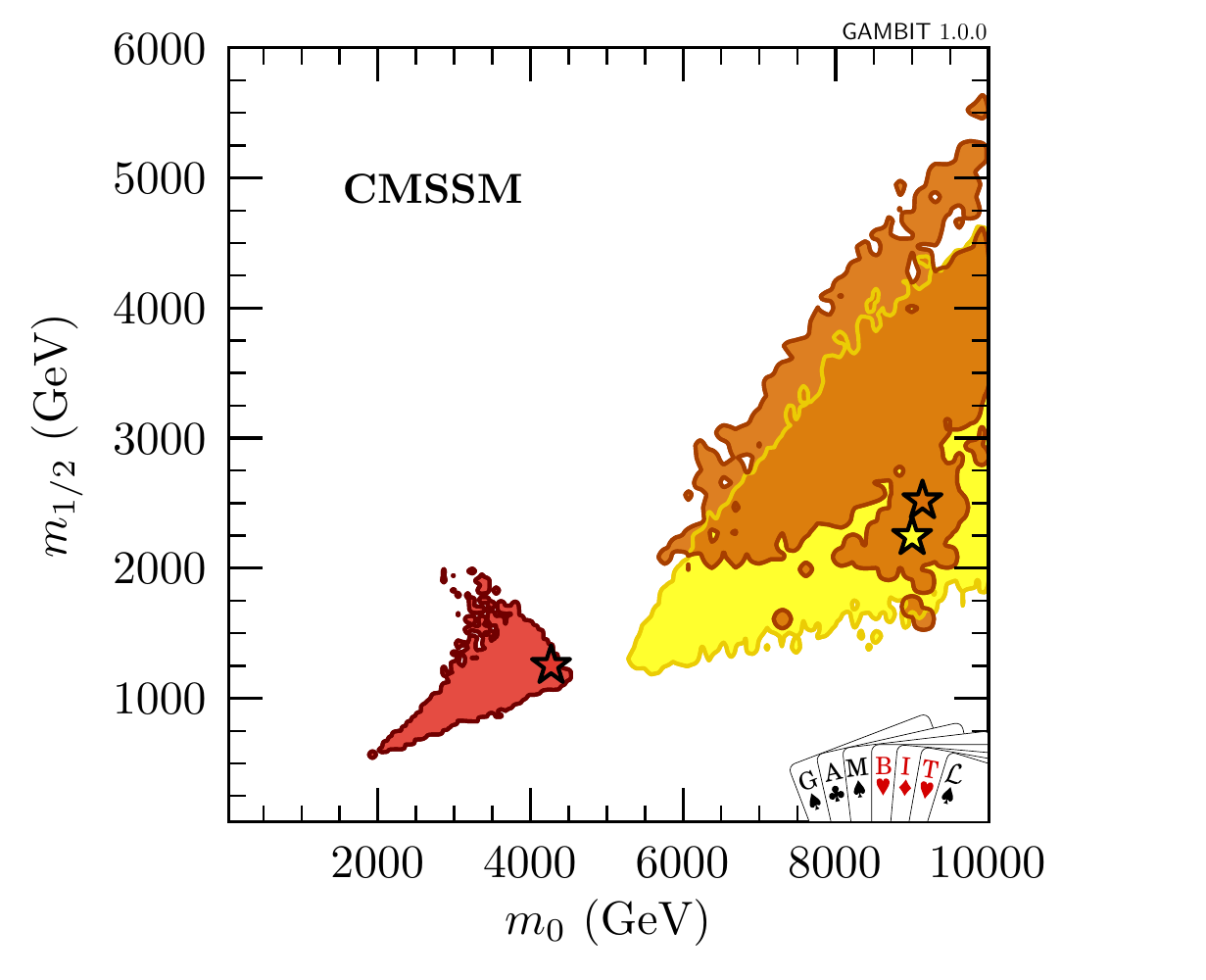}
  \includegraphics[width=0.245\textwidth]{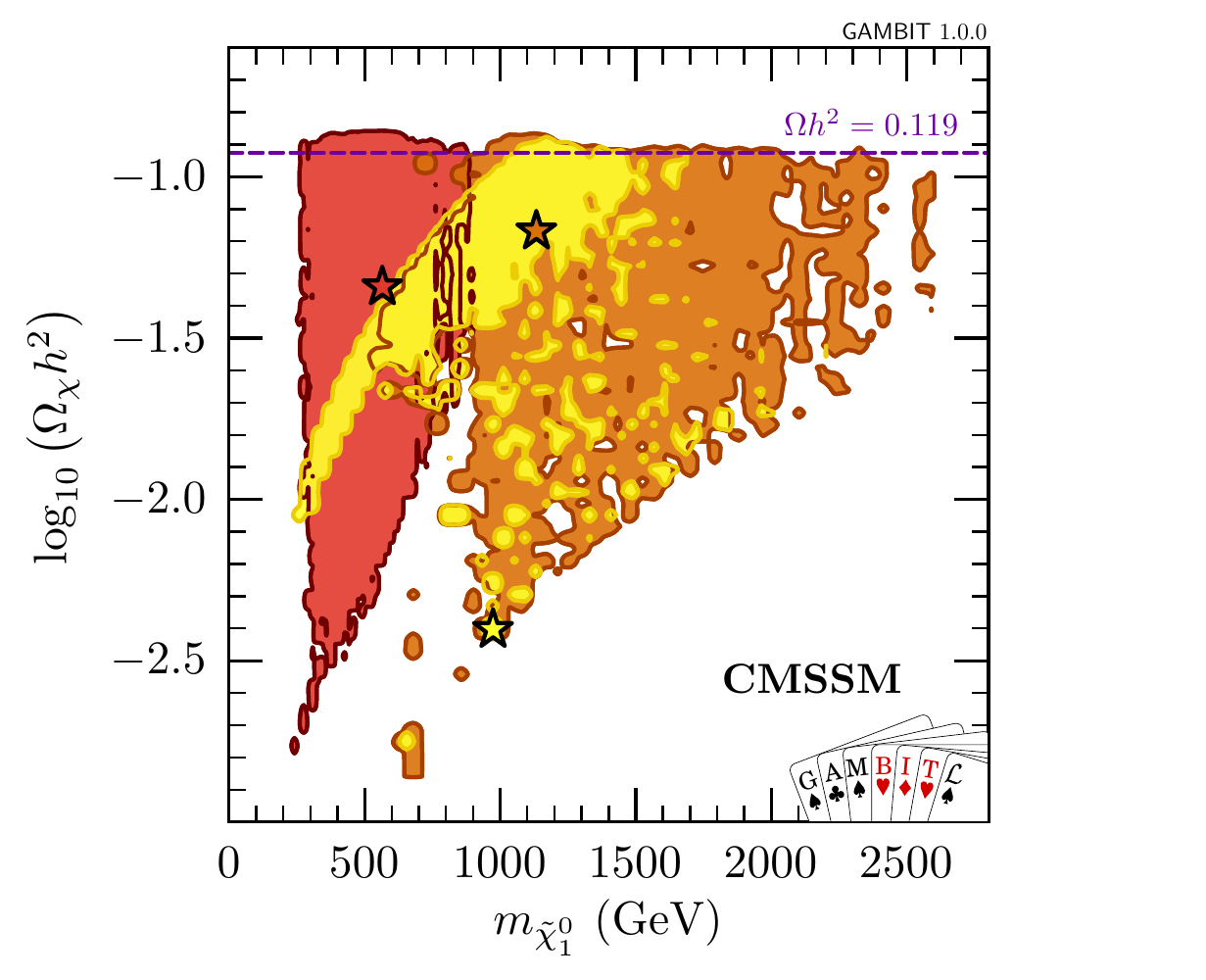}
  \includegraphics[width=0.245\textwidth]{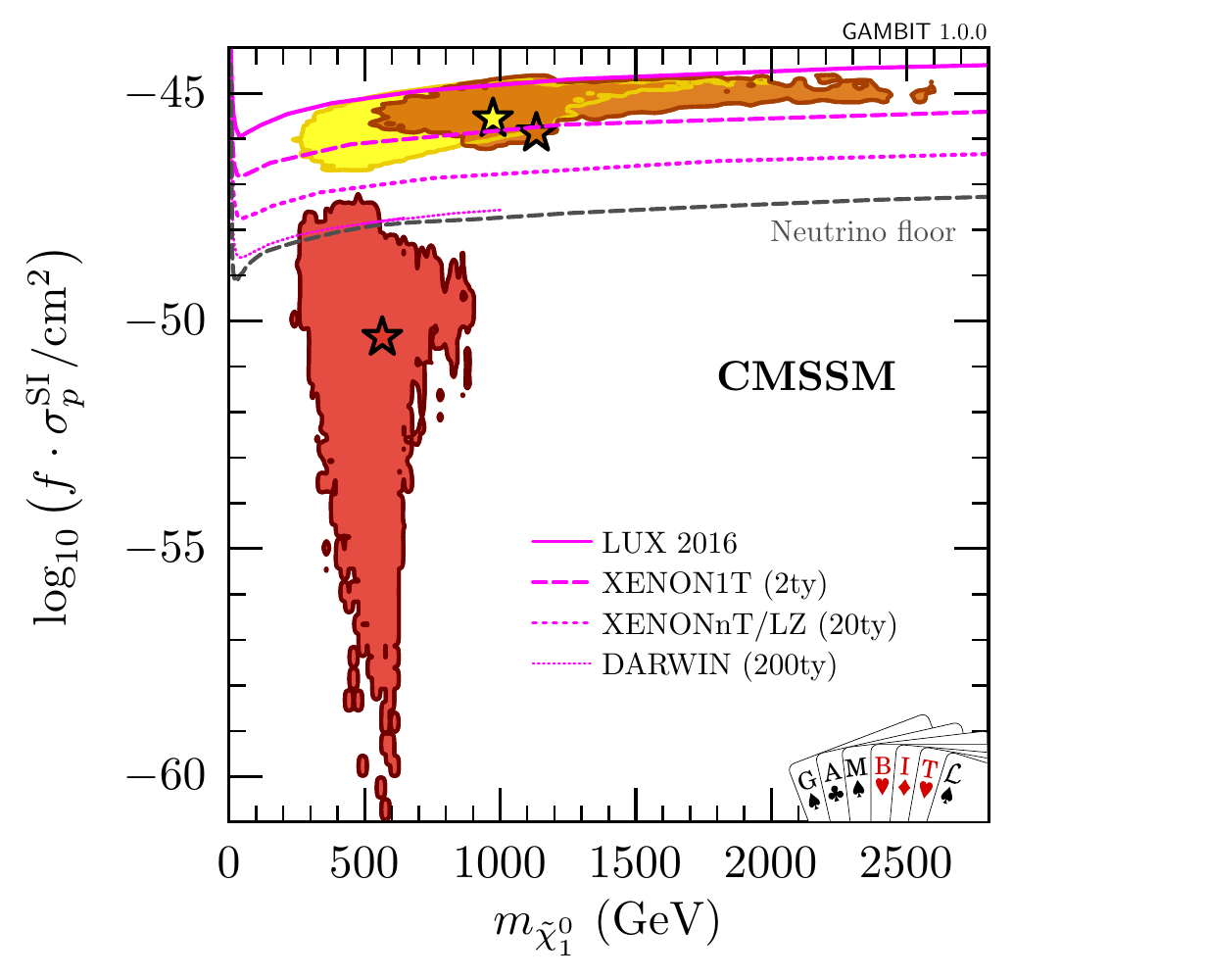}\\
 \includegraphics[height=3.77mm]{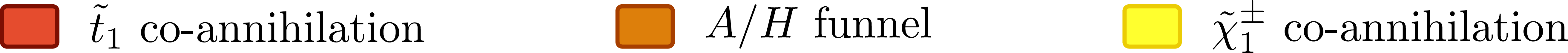}
  \caption{CMSSM results in the $m_0-m_{1/2}$ plane (left two panels), the relic abundance $\Omega_\chi h^2$ (third panel) and the spin-independent nuclear scattering cross-section (right panel).  The left panel shows the profile likelihood ratio as colour contours, with white contour lines indicating the 68\% and 95\% confidence level (CL) regions, and a star showing the best-fit point. Other panels show the mechanisms for depleting the relic density of dark matter to or below the measured value, within the 95\% CL profile likelihood regions.  Cross-sections shown in the rightmost plot are all correctly rescaled according to the predicted fraction of the relic density in neutralinos ($f$). Taken from Ref.~\cite{CMSSM}.}
  \label{fig:2d_parameter_plots_cmssm}
\end{figure*}

The results of the CMSSM global fit are shown in the $m_0-m_{1/2}$ plane and in terms of the relic density and spin-independent nuclear scattering cross-section in Fig.\ \ref{fig:2d_parameter_plots_cmssm}.  In this figure, one can see three active mechanisms to avoid DM overabundance: stop co-annihilation, chargino co-annihilation and resonant annihilation through a heavy Higgs funnel. Note that stau co-annihilation, which has previously appeared in CMSSM global fits, is now ruled out at the $95 \%$ confidence level (CL). The overall best fit point lies in the stop co-annihilation region and has stop and neutralino masses of around $600$\,GeV.  We apply the relic density constraint as an upper bound only, as it is possible that a fraction of DM may exist as another state (e.g.\ axions).  As expected, Higgsino-dominated neutralinos saturate the relic density through chargino co-annihilation at around $1$ TeV.  Hybrid Higgsino co-annihilation and heavy Higgs funnel annihilation also occurs.  The rightmost panel of Fig.\ \ref{fig:2d_parameter_plots_cmssm} shows that the chargino co-annihilation and Higgs-funnel regions can be fully probed by future direct detection experiments.  The stop co-annihilation region is very difficult to probe with direct detection, indirect detection and even the LHC, but probing the low-mass end should be possible at the LHC.  The very small direct detection cross-sections are due to fine-tuned cancellations in tree-level matrix elements.

\begin{figure*}[tbh]
  \centering
  \includegraphics[width=0.245\textwidth]{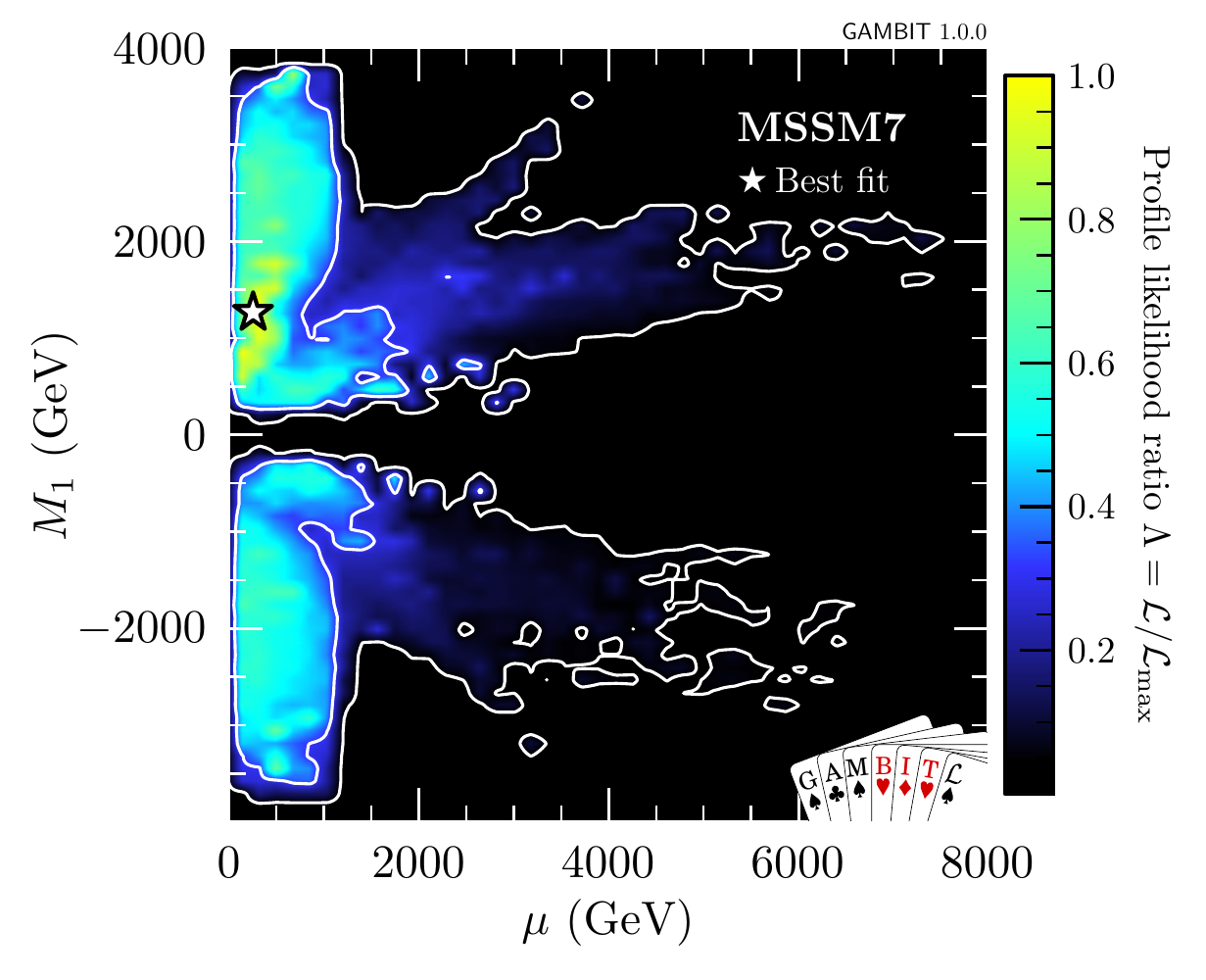}
  \includegraphics[width=0.245\textwidth]{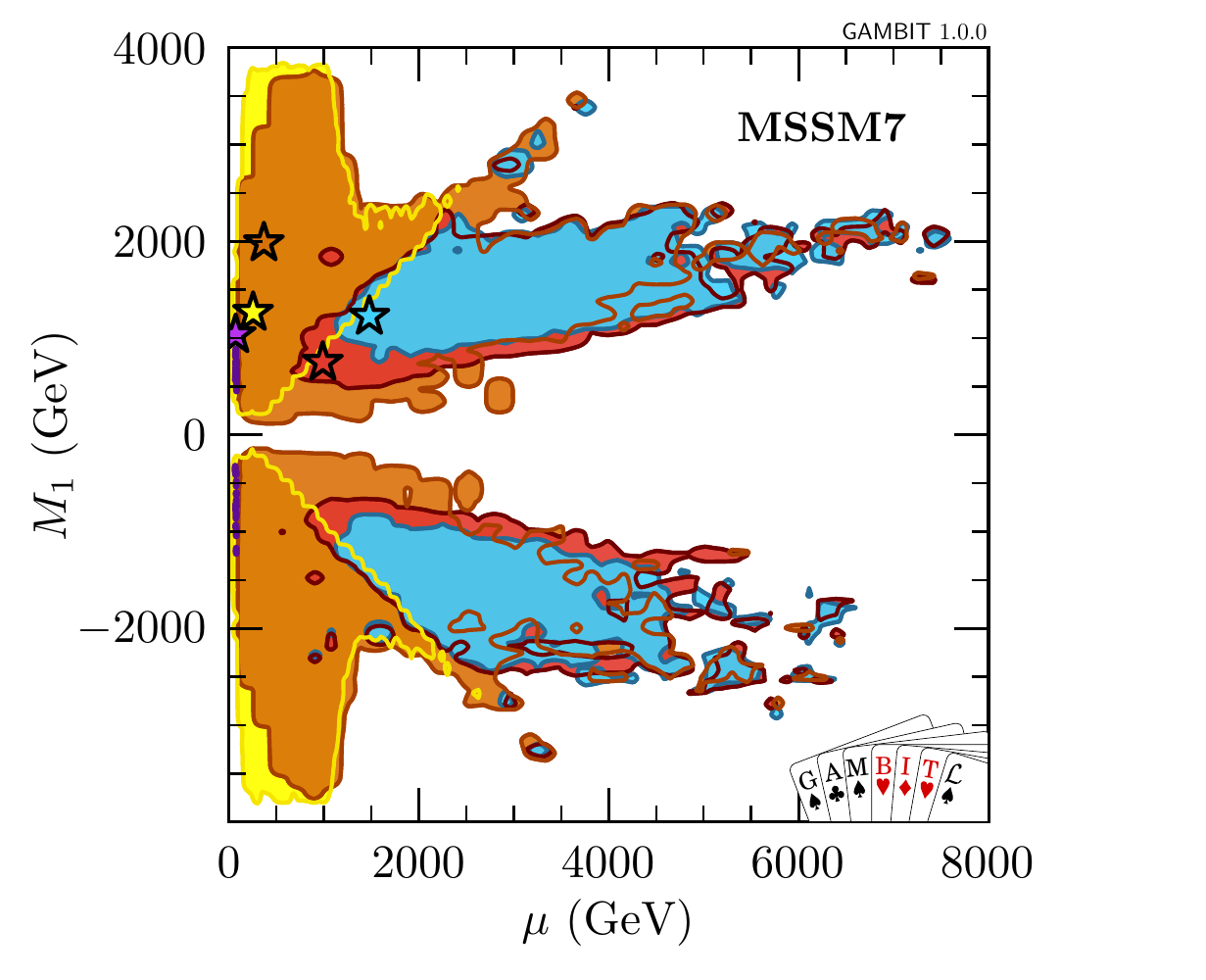}
  \includegraphics[width=0.245\textwidth]{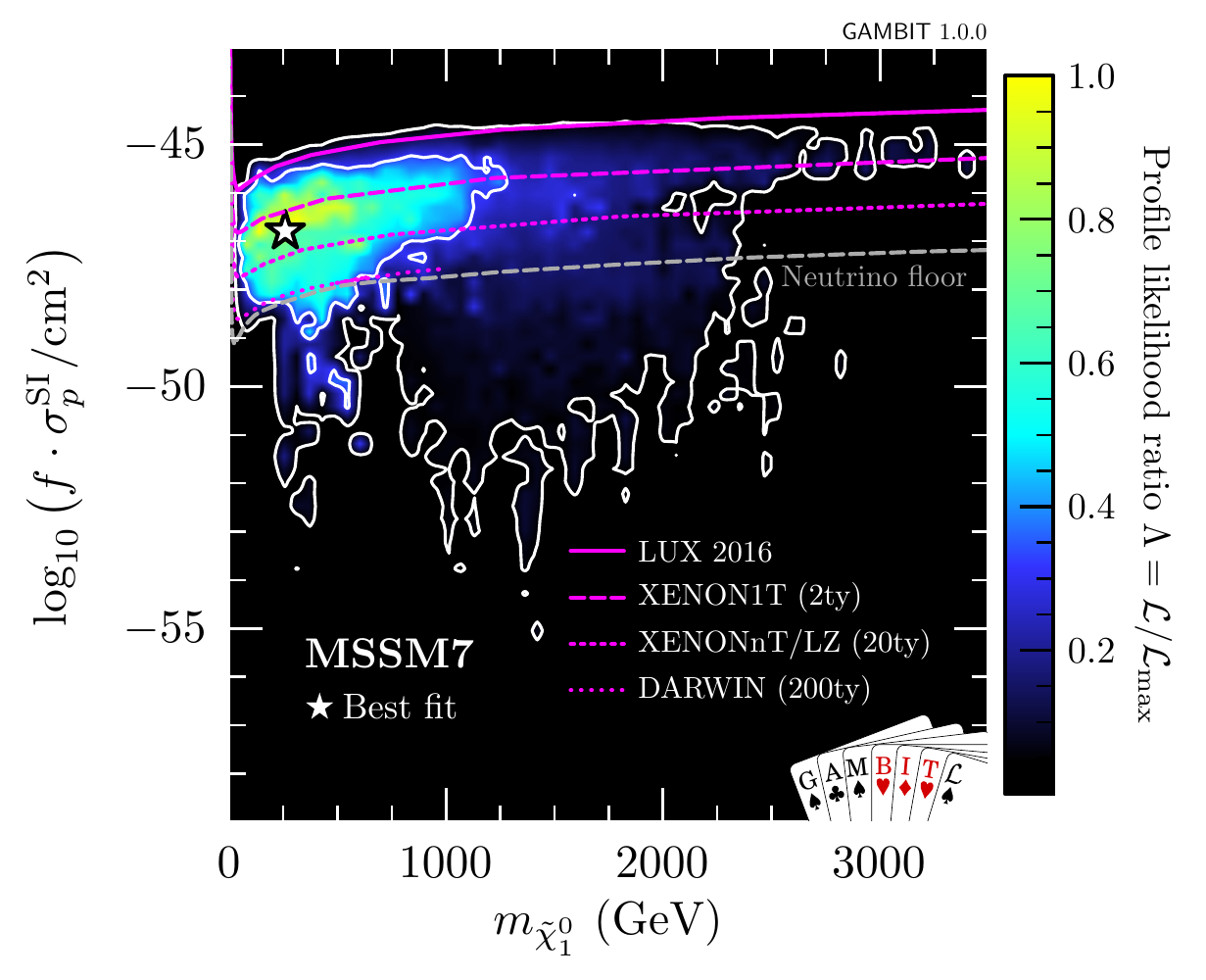}
  \includegraphics[width=0.245\textwidth]{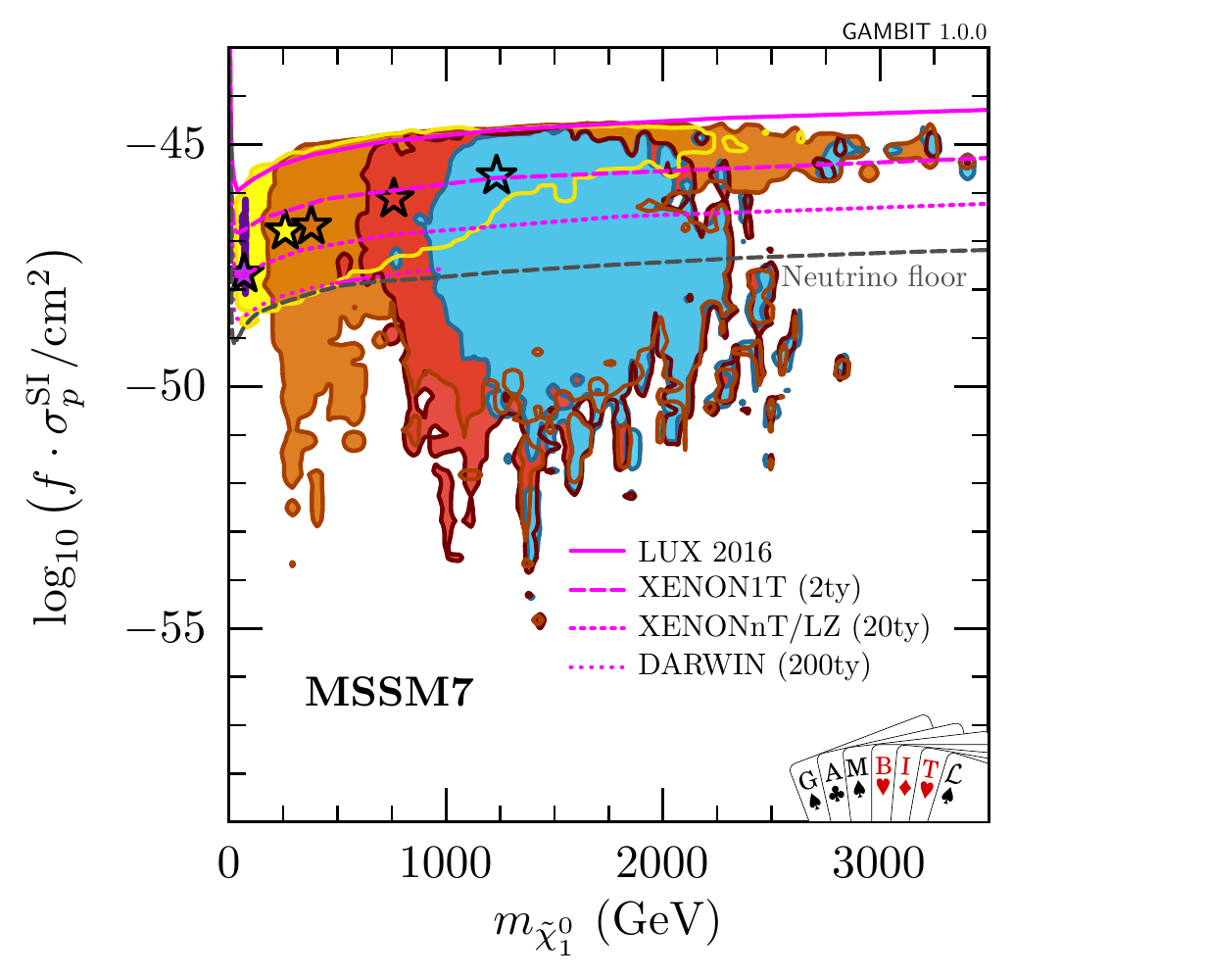}\\
  \includegraphics[height=3.77mm]{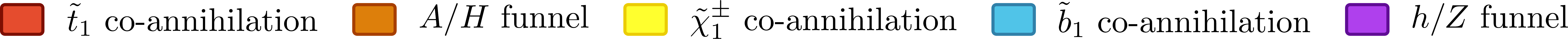}
  \caption{MSSM7 results in the $\mu$--$M_1$ plane (left two panels) and in terms of the spin-independent neutralino-proton cross-section, rescaled by the predicted relic density fraction $f$. Colours and annotations have equivalent meanings to those in Fig.\ \protect\ref{fig:2d_parameter_plots_cmssm}.  Taken from Ref.\ \cite{MSSM}.}
  \label{fig:2d_param_planes}
\end{figure*}

The left two panels of Fig.\ \ref{fig:2d_param_planes} show MSSM7 results in the $\mu$--$M_1$ plane. This is particularly useful for understanding the different lightest neutralino scenarios. We find lightest neutralinos that are Higgsino dominated, bino-dominated and a mixture of the two. Winos cannot dominate the lightest neutralino, since $M_2 \approx 2 M_1$ in the MSSM7.  The right two panels show the spin-independent nuclear scattering cross-section. The best-fit point is a chargino co-annihilation scenario with chargino and neutralino masses around $260$\,GeV and a mass difference of around $10$\,GeV.  This is challenging to detect at the LHC. However, the entire chargino co-annihilation and light Higgs funnel regions will be probed by future direct detection experiments.

\begin{figure*}[tb]
  \centering
  \includegraphics[width=0.48\textwidth]{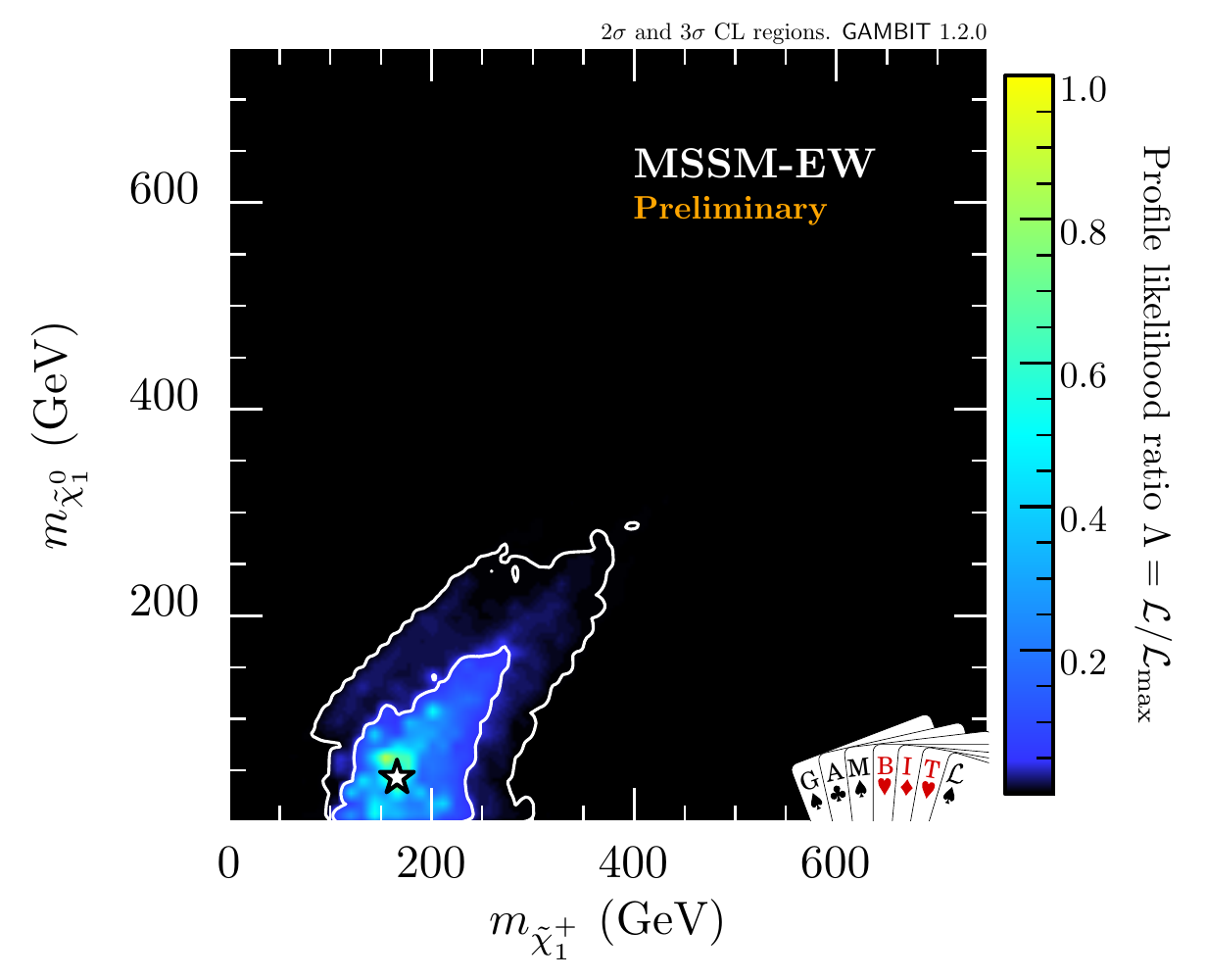}
  \includegraphics[width=0.48\textwidth]{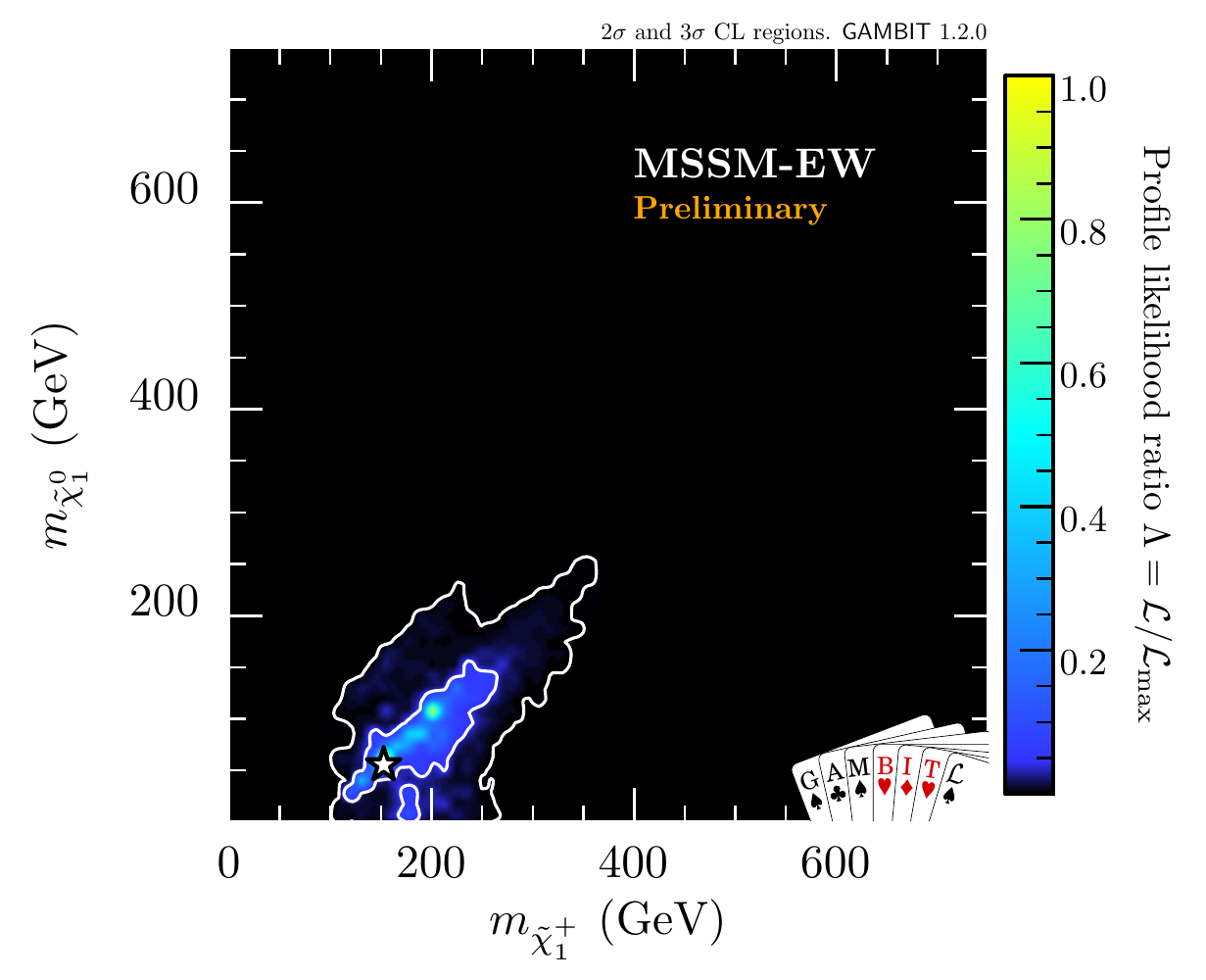}
  \caption{Preliminary results from the electroweakino scan shown in the $m_{\tilde{\chi}_1^0}-m_{\tilde{\chi}_1^\pm}$ plane. In the left panel only one (\cite{Aaboud:2018sua}) of the two ATLAS searches in two- and three-lepton final states is included in the total likelihood function, while in the right panel both searches have been included as independent likelihood contributions. White contours correspond to 95\% and 99.7\% CL regions.}
  \label{fig:EWino}
\end{figure*}

Finally, we show preliminary results from a global fit focused on the electroweakino sector of the MSSM in Fig.\ \ref{fig:EWino}. We scan only over the parameters relevant to this sector ($M_1$, $M_2$, $\mu$ and $\tan\beta$), decoupling all SUSY states other than the neutralinos and charginos. We focus our investigation on collider searches for these states, with the aim of identifying what impact these searches, which are typically interpreted in simplified models, have had on the general MSSM elektroweakino sector. Here we include the rigorous \colliderbit implementation of almost all relevant LHC searches for promptly decaying electroweakinos, using 36~fb$^{-1}$ of 13\,TeV proton-proton collision data.  This includes ATLAS searches for chargino and neutralino production in two- and three-lepton final states~\cite{Aaboud:2018jiw}, for chargino and neutralino production using recursive jigsaw reconstruction in final states with two or three leptons~\cite{Aaboud:2018sua}, for pair production of Higgsinos in the $hh$ final state~\cite{Aaboud:2018htj}, and for supersymmetry in final states with four or more leptons~\cite{Aaboud:2018zeb}.  It also includes CMS searches for chargino and neutralino production in the $Wh$ final state~\cite{CMS:2017fth}, for degenerate charginos and neutralinos in final states with two low-momentum opposite-sign leptons~\cite{CMS:2017fij}, in the ``EW'' signal regions requiring an on-shell $Z$ boson in the search for states with jets and two opposite-sign same-flavour leptons~\cite{Sirunyan:2017qaj}, and for chargino and neutralino production in final states with two or three leptons~\cite{CMS-PAS-SUS-16-039}.
For our implementation of the searches in~\cite{CMS:2017fij} and~\cite{Sirunyan:2017qaj} we follow the ``simplified likelihood'' approach introduced in~\cite{Collaboration:2242860} to make use of the approximate background covariance matrices provided by CMS.
The two ATLAS searches in ~\cite{Aaboud:2018jiw} and ~\cite{Aaboud:2018sua} both target final states with two or three leptons. However, the overlap between the final datasets for the two analyses is relatively small~\cite{Aaboud:2018sua}. In Fig.~\ref{fig:EWino} we show the fit results that we obtain when we only include the recursive jigsaw analysis from ~\cite{Aaboud:2018sua} (left panel) and when both analyses are included as independent likelihood contributions (right panel).  In addition to the LHC searches, our fit takes into account constraints on chargino and neutralino production cross-sections from searches at LEP, and the limits on the invisible decay widths of the $Z$ and the $125$\,GeV Higgs.

The preliminary result in Fig.~\ref{fig:EWino} is very intriguing, as we see closed 3$\sigma$ CL contours, suggesting the presence of a signal. The result should be interpreted with great care until the overall goodness of fit of the highlighted models has been studied in detail.

\GB results shed new light on the MSSM, giving a rigorous picture of its status in light of the many experimental constraints, and throwing up an intriguing new feature in the electroweakino sector.

\acknowledgments I gratefully acknowledge the rest of the GAMBIT Collaboration, as well as Matthias Danninger, Rose Kudzman-Blais, Andreas Petridis, Abhishek Sharma and Yang Zhang for contributions to the work presented here. I also acknowledge PRACE for awarding \GB access to Marconi at CINECA, Italy.

%\begin{thebibliography}{99}
%\bibitem{...} ....
%\end{thebibliography}

\bibliographystyle{JHEP_pat}
\bibliography{GAMBIT_MSSM}

\end{document}